\documentclass[a4paper,12pt]{article}
\usepackage{graphicx}
\usepackage{amsmath}
\usepackage{hyperref}
\usepackage{xcolor}
\usepackage{enumerate}
\title{Exploring the transition between Quantum and Classical
Mechanics}

\author{E. Aldo Arroyo\thanks{aldo.arroyo@ufabc.edu.br},\\
    Centro de Ci\^{e}ncias Naturais e Humanas, Universidade Federal do ABC,\\
    Santo Andr\'{e}, 09210-170 S\~{a}o Paulo, SP, Brazil.}

\date{\today}

\begin{document}
    \maketitle
    \begin{abstract}
We investigate the transition from quantum to classical mechanics
using a one-dimensional free particle model. In the classical
analysis, we consider the initial positions and velocities of the
particle drawn from Gaussian distributions. Since the final
position of the particle depends on these initial conditions,
convolving the Gaussian distributions associated with these
initial conditions gives us the distribution of the final
positions. In the quantum scenario, using an initial Gaussian wave
packet, the temporal evolution provides the final wave function,
and from it, the quantum probability density. We find that the
quantum probability density coincides with the classical normal
distribution of the particle's final position obtained from the
convolution theorem. However, for superpositions of Gaussian
distributions, the classical and quantum results deviate due to
quantum interference. To address this issue, we propose a novel
approach to recover the classical distribution from the quantum
one. This approach involves removing the quantum interference
effects through truncated Fourier analysis. These results are
consistent with modern quantum decoherence theory. This
comprehensive analysis enhances our understanding of the
classical-quantum correspondence and the mechanisms underlying the
emergence of classicality from quantum systems.
    \end{abstract}

\tableofcontents

\section{Introduction}

The transition from quantum to classical behavior remains a
central question in physics. While classical mechanics excels in
describing macroscopic systems, the precise mechanisms by which
classicality emerges from the underlying quantum realm are not
fully elucidated, with decoherence being one of the currently
known phenomena in this field \cite{Wojciech1,Wojciech2}.
Decoherence poses a significant challenge in advancing quantum
information technologies, including the realization of practical
quantum computers \cite{Buluta1}.

In this work, we investigate the transition from quantum to
classical mechanics by considering the example of a
one-dimensional free particle \cite{Maamache1,bagrov2014coherent}.
In the classical analysis, the initial positions $x_i$ and
velocities $v_i$ of the particle are randomly drawn from Gaussian
distributions. Since at a given instant $t$, the final position
$x_f$ of the particle depends on the initial conditions $x_i$ and
$v_i$, the distribution of $x_f$ as a function of time can be
obtained by convolving the distributions of $x_i$ and $v_i$. In
the quantum case, the initial wave function $\Psi(x,0)$ is chosen
as a Gaussian. The temporal evolution of this wave function
through the Schr\"{o}dinger equation provides the final wave
function $\Psi(x,t)$, from which the quantum probability density
$|\Psi(x,t)|^2$ can be calculated. It turns out that this quantum
probability density coincides with the normal distribution of
$x_f$ obtained in the classical analysis.

To our knowledge, this specific result has not been addressed in
the existing literature, suggesting its relevance and importance
in understanding the transition from quantum to classical
mechanics. It is worth noting that in quantum mechanics, the
classical limits of quantum systems have been studied, where
classical properties emerge approximately from the quantum
description, in a concept known as the correspondence principle
\cite{corres1,corres2,ballentine1994inadequacy}.

Initially, the discovery that classical and quantum probability
densities turn out to be equivalent in the specific case studied
in this work suggests a fascinating possibility: that quantum
systems could potentially be described from classical mechanics by
considering random initial conditions. However, as we will see in
the model of the free particle, by exploring how these
distributions evolve for a superposition of two Gaussians, i.e.,
considering a scenario analogous to the double-slit experiment
\cite{arndt1999wave,FORD200187,Gobert1}, we find that classical
and quantum descriptions diverge when the initial state is a
superposition due to quantum interference. In this case, we
propose a novel approach to recover the classical distribution
from the quantum one using truncated Fourier transform analysis.
This method effectively removes high-frequency components
associated with quantum interference, leading to a distribution
similar to the classical one. To our knowledge, this application
of truncated Fourier analysis in this context is a novel
contribution to the field. It is important to note that while our
study in the classical regime involves random initial conditions,
it is distinct from the stochastic interpretation of quantum
mechanics \cite{lindgrenquantum1}, as we maintain the
deterministic nature of the classical equations of motion.

This study illuminates the connection between classical and
quantum probability distributions through the introduction of a
novel method employing Fourier analysis. It emphasizes the
significance of quantum interference in delineating the boundaries
between classical and quantum regimes. These findings contribute
to a deeper understanding of how classical behavior emerges from
the underlying quantum world.

This paper is structured as follows. In Section 2, we analyze the
classical free particle with random initial conditions and compute
the distributions of its final position using the convolution
theorem. We then compare this result with the quantum probability
density. In Section 3, we study the superposition of two Gaussian
distributions and compare the classical and quantum results. In
Section 4, we recover the classical probability density by
truncating the Fourier components of the quantum probability
density. In Section 5, for completeness, we test the consistency
of our results in the context of quantum decoherence theory.
Section 6 provides a summary of the findings and discusses further
directions for exploration.

\section{Free particle with random initial conditions}
Let us analyze the scenario in which the initial position,
$x_{i}$, and initial velocity, $v_{i}$, of a free particle with
mass $m$ moving along the $x$-axis are randomly drawn from normal
Gaussian distributions. Specifically, these Gaussian distributions
are represented by the following probability density functions
(PDFs):

\begin{align}
\label{fXx} P_{X_i}(x) &= \frac{1}{\sqrt{2\pi \sigma_x^2}}
e^{-\frac{(x - x_0)^2}{2\sigma_x^2}}, \\
\label{fVv} P_{V_i}(v) &= \frac{1}{\sqrt{2\pi \sigma_v^2}}
e^{-\frac{(v - v_0)^2}{2\sigma_v^2}}.
\end{align}
Here, $\sigma_x$ and $\sigma_v$ represent the standard deviations,
while the mean values, $x_0$ and $v_0$, signify the most likely
values to appear as the initial position and velocity of the
particle.

For a given initial position $x_{i}$ and initial velocity $v_{i}$
at some time $t$, we can calculate the corresponding final
position $x_{f}$ using the equation $x_{f} = x_{i} + v_{i} t$. By
repeating this process multiple times and generating new sets of
initial positions and velocities, we obtain a collection of
different final positions $x_{f}$ for each repetition. Our
objective is to determine the normal distribution of final
positions $x_{f}$ at any given time $t$. To determine this normal
distribution, we can utilize the convolution theorem.

According to the convolution theorem, the probability density
function (PDF) of $x_f$, denoted as $P_{X_f}(x,t)$, can be
obtained by convolving the PDFs of $x_{i}$ and $v_{i}$ using the
integral:
\begin{align}
\label{fXfx} P_{X_f}(x,t) = \int_{-\infty}^{\infty}
P_{X_i}(x-vt)P_{V_i}(v) \, dv.
\end{align}

By substituting the PDFs, $P_{X_i}(x)$ and $P_{V_i}(v)$, given in
equations (\ref{fXx}) and (\ref{fVv}), into the convolution
formula (\ref{fXfx}), we can perform the integration and obtain
the following expression:
\begin{align}
\label{fXfx2} P_{X_f}(x,t) =\frac{1}{\sqrt{2\pi (\sigma_x^2 +
\sigma_v^2 t^2)}}  \,  \text{exp} \Big\{-\frac{\big(x - (x_0 + v_0
t)\big)^2}{2(\sigma_x^2 + \sigma_v^2  t^2)}\Big\}.
\end{align}

This formula (\ref{fXfx2}) represents the normal distribution of
the final positions at any given time $t$, considering the random
selection of initial positions and velocities from normal Gaussian
distributions. Notably, the mean of the final position, $x_0 + v_0
t$, aligns with our expectations. Furthermore, the standard
deviation of the distribution is determined by $\sqrt{\sigma_x^2 +
\sigma_v^2 t^2}$. This indicates that the spread of the
distribution increases with time, reflecting the accumulated
effects of the initial position and velocity uncertainties.

Now let us consider, in a one-dimensional model, the quantum free
particle whose initial wave function at time $t=0$ is given by
\begin{align}
\label{psiini1} \Psi(x,0) = \Big( \frac{1}{2\pi \sigma^2}
\Big)^{1/4} e^{i m v_0 x/\hbar}\, e^{-(x-x_0)^2/4\sigma^2 }.
\end{align}
The normalized wave function given by equation (\ref{psiini1})
exhibits a peak at $x_0$, indicating the particle's most probable
position. The parameter $v_0$ corresponds to the group velocity of
the wave packet. The parameter $\sigma$ determines the width of
the Gaussian.

To calculate the wave function $\Psi(x,t)$ at time $t$, we can use
the following formula:
\begin{align}
 \label{psiinit1} \Psi(x,t) = \sqrt{\frac{m}{2 \pi \hbar \, i \, t}}
 \; \int_{-\infty}^{\infty} \Psi(y,0) \; e^{im(x-y)^2/(2 \hbar \, t)} \;
 dy.
\end{align}
By substituting equation (\ref{psiini1}) into equation
(\ref{psiinit1}) and performing the Gaussian integrals, we obtain
a lengthy expression for $\Psi(x,t)$. However, our primary
interest lies in the square of the wave function, which
corresponds to the probability density of the particle at time
$t$. This probability density is given by
\begin{align}
 \label{psiinident1} |\Psi(x,t)|^2 =  \frac{1}{\sqrt{2\pi(\sigma^2+\frac{\hbar^2 t^2}{4 \sigma^2 m^2}})}
 \exp \left(-\frac{\left(x-(x_0+v_0 t)\right)^2}{2(\sigma^2+\frac{\hbar^2 t^2}{4\sigma^2 m^2})}\right).
\end{align}

At this point, let us show a remarkable correspondence between the
classical normal distribution of particle positions, as
represented by the PDF given in equation (\ref{fXfx2}), and the
quantum probability density described in equation
(\ref{psiinident1}) associated with the temporal evolution of a
free quantum particle's wave function. By making the following
identifications:
\begin{align}
 \label{correspon1} \sigma_x \rightarrow \sigma, \;\;\; \text{and} \;\;\; \sigma_v \rightarrow
 \frac{\hbar}{2 \sigma m },
\end{align}
equation (\ref{fXfx2}) precisely aligns with equation
(\ref{psiinident1}).

This result has significant implications for our understanding of
quantum mechanics and its relationship to classical physics.

By discovering that the probability density of a free particle,
with random initial conditions, in classical mechanics can be
described by a similar Gaussian form to that of a quantum
particle's wave function, we establish a profound connection
between classical and quantum descriptions. Specifically, using
the identifications (\ref{correspon1}), we find that the quantum
probability density can be reproduced by the classical result
obtained through the convolution theorem.

However, it is important to note that the results from classical
mechanics fall short in capturing quantum phenomena such as
interference. While the convolution of classical Gaussian
distributions does not produce interference patterns, the quantum
mechanical description inherently includes interference effects
due to the superposition principle. In subsequent studies, we will
explore the interference phenomena by superposing quantum Gaussian
wave packets.

Moreover, we will investigate the transition from classical to
quantum behavior, specifically how classical Gaussian
distributions, when superposed, do not produce interference
patterns, while quantum mechanical descriptions do. To understand
this transition, we will consider the limit where $\hbar
\rightarrow 0$. However, as we will see, this limit alone is
insufficient to eliminate the interference patterns.

\section{Superposition of Gaussian distributions: A classical and quantum comparison}
In this section, we aim to study the superposition of two
Gaussians using classical and then quantum mechanics. In the
classical case, no interference patterns are observed, whereas in
the quantum case, interference patterns do appear. Furthermore,
when taking the limit \(\hbar \rightarrow 0\), the quantum result
does not necessarily tend to or approximate the classical result.
This indicates that the limit \(\hbar \rightarrow 0\) does not
always correspond to classical mechanics in the sense that taking
the limit \(c \rightarrow \infty\) in special relativity recovers
Newtonian mechanics.

To study the superposition of two Gaussians classically, we
consider the initial position distribution \(P_{X_i}(x)\) as a sum
of two Gaussians:
\begin{align}
\label{psup1} P_{X_i}(x) = \frac{1}{2 \sqrt{2 \pi \sigma^2}}
\left[ e^{-\frac{(x - \frac{d}{2})^2}{2\sigma^2}} + e^{-\frac{(x +
\frac{d}{2})^2}{2\sigma^2}} \right],
\end{align}
where \(d\) is the distance between the peaks of the Gaussians,
and \(\sigma\) is the parameter characterizing the width of each
Gaussian. This implies that the initial position \(x_i\) has the
highest probability around \(+d/2\) for \(x > 0\) or \(-d/2\) for
\(x < 0\).

For the initial velocities, to ensure that the two Gaussians move
towards each other as time progresses ($t>0$), we require the
Gaussian centered at \(x = -d/2\) to move to the right, implying
initial velocities \(v_i> 0\) for \(x_i < 0\). Similarly, the
Gaussian centered at \(x = d/2\) moves to the left, implying
initial velocities \(v_i < 0\) for \(x_i > 0\). These initial
velocities are also drawn from Gaussian distributions:
\begin{align}
\label{psup2}
P_{V_{i+}}(v) = \frac{1}{\sqrt{2\pi \sigma_v^2}} e^{-\frac{(v - v_0)^2}{2\sigma_v^2}} \quad \text{for} \quad x_i < 0 \quad \text{and} \quad v_i > 0,\\
\label{psup3} P_{V_{i-}}(v) = \frac{1}{\sqrt{2\pi \sigma_v^2}}
e^{-\frac{(v + v_0)^2}{2\sigma_v^2}} \quad \text{for} \quad x_i >
0 \quad \text{and} \quad v_i < 0.
\end{align}

Using these distributions (\ref{psup1}), (\ref{psup2}), and
(\ref{psup3}) for the initial position and velocity, we calculate
the distribution of the final positions \(x_f = x_i + v_i t\).
Applying the convolution theorem, we obtain the following
classical probability density, or simply classical distribution:
\begin{align}
\label{psup4} P_{\text{Cl}}(x, t) = \frac{1}{2 \sqrt{2 \pi
(\sigma^2 + \sigma_v^2 t^2)}} \left[ e^{-\frac{(x - \frac{d}{2} +
v_0 t)^2}{2 (\sigma^2 + \sigma_v^2 t^2)}} + e^{-\frac{(x +
\frac{d}{2} - v_0 t)^2}{2 (\sigma^2 + \sigma_v^2 t^2)}} \right].
\end{align}

As we can observe from this result, the total Gaussian
distribution is simply the sum of two individual Gaussians that
were initially centered at $x=-d/2$ and $x=d/2$ and then move
towards each other with an average velocity $v_0$. Consequently,
no interference patterns are observed in this classical
superposition.

To study the same problem from the quantum perspective, we will
consider the following initial wave function:
\begin{align}
\label{psup5} \Psi(x,0) =
\frac{e^{-\frac{\left(x-\frac{d}{2}\right)^2}{4 \sigma^2}-\frac{i
m v_0 x}{\hbar}}+e^{-\frac{\left(\frac{d}{2}+x\right)^2}{4
\sigma^2}+\frac{i m v_0 x}{\hbar}}}{(2\pi)^{1/4}\sqrt{2\sigma}
\sqrt{1+e^{-\frac{d^2}{8 \sigma^2}-\frac{2 m^2 \sigma^2
v_0^2}{\hbar^2}}}},
\end{align}
which, similar to the classical case, represents the superposition
of two Gaussian wave packets with peaks centered at \(x=d/2\) for
\(x>0\), and the peak centered at \(x=-d/2\) for \(x<0\). As time
progresses (\(t>0\)), we will observe the Gaussian with the peak
at \(x=-d/2\) moving to the right with a positive velocity
\(v_0>0\), and the peak at \(x=d/2\) moving to the left with a
velocity of \(-v_0<0\).

Calculating the square of the initial wave function (\ref{psup5}),
we obtain:
\begin{align}
\label{psup6} |\Psi(x,0)|^2 = \frac{2 e^{-\frac{d^2+4 x^2}{8
\sigma ^2}} \cos \left(\frac{2 m v_0
x}{\hbar}\right)+e^{-\frac{\left(x-\frac{d}{2}\right)^2}{2 \sigma
^2}}+e^{-\frac{\left(\frac{d}{2}+x\right)^2}{2 \sigma ^2}}}{2
\sqrt{2 \pi \sigma^2}  \left(1+e^{-\frac{d^2}{8 \sigma ^2}-\frac{2
m^2 \sigma ^2 v_0^2}{\hbar^2}}\right)}.
\end{align}

To compare the classical distribution of initial particle
positions, given by equation (\ref{psup1}), with the square of the
initial wave function in the quantum case, represented by equation
(\ref{psup6}), we observe that they are not equal. However, by
considering the following conditions: $\frac{d}{\sigma}$ is
sufficiently large, specifically $\frac{d}{\sigma} \gg 1$, and
$\frac{m \sigma v_0}{\hbar} \gg 1$, we find that the terms
$e^{-\frac{d^2}{8 \sigma^2}-\frac{2 m^2 \sigma^2 v_0^2}{\hbar^2}}$
and $e^{-\frac{d^2+4 x^2}{8 \sigma^2}} \cos \left(\frac{2 m v_0
x}{\hbar}\right)$ vanish. This implies that the initial
distributions (\ref{psup1}) and (\ref{psup6}) are equal under
these conditions.

To clarify, under these conditions $\frac{d}{\sigma} \gg 1$, and
$\frac{m \sigma v_0}{\hbar} \gg 1$, for $t=0$, we can ensure that
both the classical initial position distribution (\ref{psup1}) and
the quantum initial probability density (\ref{psup6}) are
practically equal. Next, we will investigate whether this equality
persists for $t>0$, namely we will proceed to study the temporal
evolution.

For the classical case, we already have the solution for the
position distribution at a time $t>0$, given by equation
(\ref{psup4}). Now, for the quantum case, using the initial wave
function given in equation (\ref{psup5}) and the formula
(\ref{psiinit1}), we can calculate the wave function $\Psi(x,t)$.
However, our primary interest lies in the square of the wave
function, which is given by:
\begin{align}
 |\Psi(x,t)|^2 =& \frac{1}{2  \sqrt{2 \pi(\sigma ^2+\frac{\hbar^2 t^2}{4 m^2 \sigma ^2})} \left(1+e^{-\frac{d^2}{8 \sigma ^2}-\frac{2
   m^2 \sigma ^2 v_0^2}{\hbar^2}}\right)} \Big[ e^{-\frac{\left(x+\frac{d}{2}-t v_0\right){}^2}{\frac{\hbar^2 t^2}{2 m^2 \sigma ^2}+2 \sigma ^2}} +
   e^{-\frac{\left(x-\frac{d}{2}+t v_0\right){}^2}{\frac{\hbar^2 t^2}{2 m^2 \sigma ^2}+2 \sigma ^2}} \nonumber
   \\ \label{psup7}
   &  + 2 \exp \left(-\frac{d^2-4 d t v_0+4 \left(t^2 v_0^2+x^2\right)}{\frac{2 \hbar^2 t^2}{m^2 \sigma ^2}+8 \sigma ^2}\right)
   \cos \left(\frac{m x \left(d \hbar^2 t+8 m^2 \sigma ^4 v_0\right)}{\hbar^3 t^2+4 \hbar m^2 \sigma ^4}\right) \Big]  .
\end{align}

Let us compare the quantum result (\ref{psup7}) with the classical
result for the position distribution at a time $t > 0$, given by
equation (\ref{psup4}). Comparing these two equations, we first
observe that $\sigma_v = \frac{\hbar}{2m\sigma}$, which is
consistent with the result shown in equation (\ref{correspon1}).
If we consider the conditions $\frac{d}{\sigma} \gg 1$ and
$\frac{m \sigma v_0}{\hbar} \gg 1$, the two expressions
(\ref{psup4}) and (\ref{psup7}) become almost identical, except
for the interference term that appears in the quantum case.

It is worth noting that, for the initial time $t=0$, using the
conditions $\frac{d}{\sigma} \gg 1$ and $\frac{m \sigma
v_0}{\hbar} \gg 1$, it was possible to eliminate the interference
term. Nevertheless, these conditions for $t > 0$ do not eliminate
the quantum interference term. As a result, the classical and
quantum solutions for $t > 0$ differ, as expected, because quantum
interference is inevitable in the quantum case.

To elucidate the results further, some graphs comparing the
solutions given by equations (\ref{psup4}) and (\ref{psup7}) will
be plotted. A graphical representation will greatly aid in
understanding the differences between the classical and quantum
cases.

For this purpose, the values $\sigma=m=1$, and $d=v_0=10$ will be
considered. By treating $\hbar$ as a free parameter, multiple
graphs with different values of $\hbar$ can be plotted to observe
the behavior as $\hbar \rightarrow 0$. The graphs will provide a
visual comparison of the classical and quantum probability density
distributions at different times.

\begin{figure}[h] \centering
    \includegraphics[width=\linewidth]{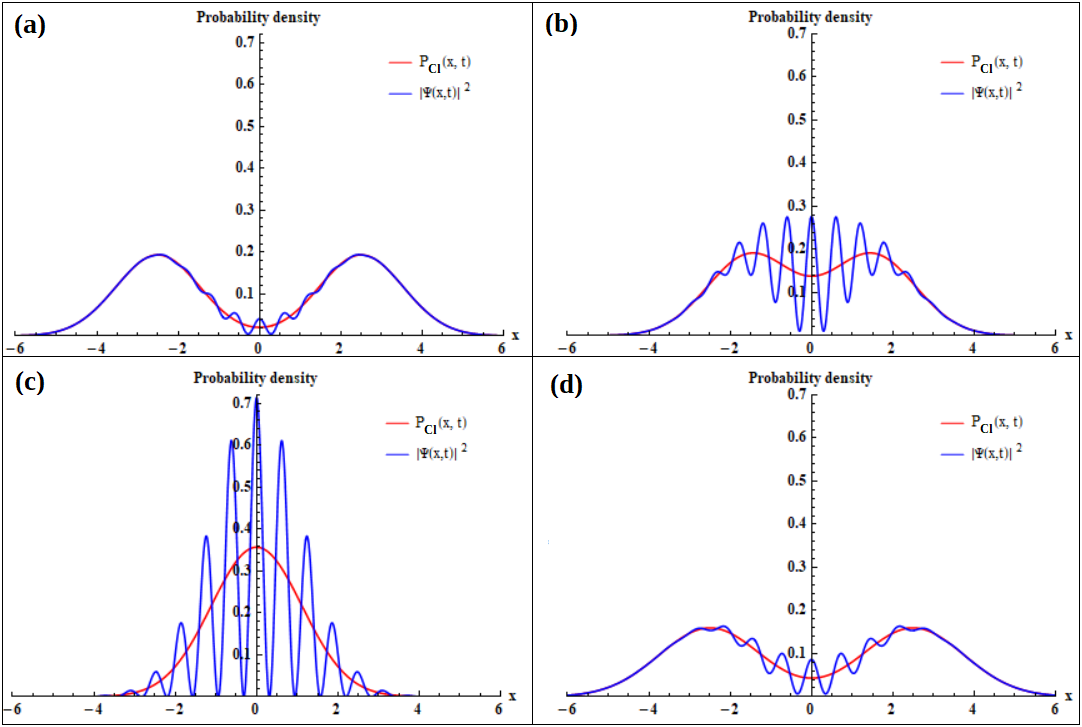}
    \caption{The following graphs depict curves defined by the functions $P_{\text{Cl}}(x, t)$ in red and $|\Psi(x,t)|^2$ in blue.
    The curves were plotted for the parameter values $\hbar = 2$, $\sigma = m = 1$, and $d = v_0 = 10$.
    The plots are presented in the order (a) $t = 0.25$, (b) $t = 0.35$, (c) $t = 0.50$, and (d) $t = 0.75$.}
    \label{fig1}
\end{figure}

In Figure \ref{fig1}, we present the classical distribution \(
P_{\text{Cl}}(x, t) \) (red) and the quantum probability density
\( |\Psi(x,t)|^2 \) (blue) for different times \( t \). The
parameters used are $\hbar = 2$, $\sigma = m = 1$, and $d = v_0 =
10$. The comparison illustrates the presence of quantum
interference as the two Gaussian wave packets approach each other.
Note that the interference terms remain significant for \( t > 0
\), manifesting as multiple oscillations with maxima and minima in
the blue curves, which are not present in the red curves
representing the classical case.

\begin{figure}[h] \centering
    \includegraphics[width=\linewidth]{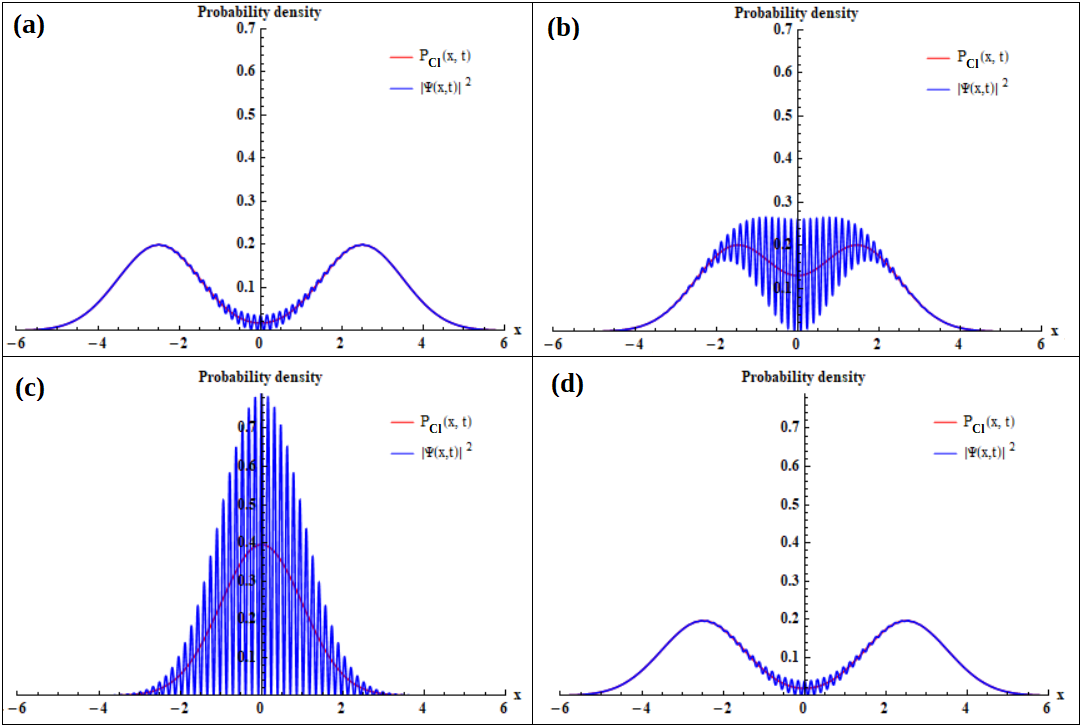}
    \caption{The following graphs depict curves defined by the functions $P_{\text{Cl}}(x, t)$ in red and $|\Psi(x,t)|^2$ in blue.
    These curves were plotted using the same parameter values as in Figure \ref{fig1}, except
    for a smaller $\hbar$ value of 0.5, compared to $\hbar = 2$ of the previous figure.
    The plots are presented in the order (a) $t = 0.25$, (b) $t = 0.35$, (c) $t = 0.50$, and (d) $t = 0.75$.}
    \label{fig2}
\end{figure}

Continuing with the analysis, in Figure \ref{fig2}, we will now
present additional plots of the same functions, \(
P_{\text{Cl}}(x, t) \) and \( |\Psi(x,t)|^2 \), using the same
parameter values \(\sigma = m = 1\) and \(d = v_0 = 10\). However,
for these new plots, we will use a significantly smaller value for
\(\hbar\). Specifically, we will set \(\hbar = 0.5\), in contrast
to the previous value of \(\hbar = 2\). This adjustment allows us
to observe the impact of a smaller \(\hbar\) on the quantum
interference patterns.

The comparison between the results shown in Figures \ref{fig1} and
\ref{fig2} reveals that reducing \(\hbar\) does not eliminate the
quantum interference effect. Instead, as \(\hbar\) is decreased,
the graph of the probability density \(|\Psi(x,t)|^2\) exhibits
increasing oscillations. This observation aligns with the
theoretical understanding that, while the limit \(\hbar \to 0\) is
often associated with the classical limit in quantum mechanics,
similar to how Newtonian mechanics is recovered from relativity in
the limit \(c \to \infty\), in this specific example, reducing
\(\hbar\) amplifies the interference patterns rather than
diminishing them.

An important observation can be made by comparing the quantum and
classical graphs of the probability density. It is evident that
the probability density \(|\Psi(x,t)|^2\), represented by the blue
curve, oscillates around the classical probability
\(P_{\text{Cl}}(x,t)\), represented by the red curve.
Conceptually, one might think of the classical value as a
baseline, with the quantum value oscillating around it, and as
\(\hbar\) decreases, these oscillations become more rapid.

This suggests that the classical value can be interpreted as an
average or mean of these quantum oscillations. This observation
led to the development of the idea of recovering the classical
result through Fourier analysis. To eliminate the interference
pattern, we will examine an approach that considers the wavenumber
components of the probability density. By filtering out certain
wavenumbers, we can effectively eliminate the interference
pattern. In the next section, we will elaborate on this approach,
providing explicit details and mathematical derivations.

\section{Recovering classical probability density by truncating Fourier components}

To obtain the classical probability density from the quantum
probability density, we begin by calculating the Fourier transform
of the quantum probability density, \(|\Psi(x,t)|^2\), as shown
below:
\begin{equation}
\label{claquan1}   \widetilde{P}(k,t) = \frac{1}{\sqrt{2
\pi}}\int_{-\infty}^{\infty} |\Psi(x,t)|^2 e^{-i k x} \, dx,
\end{equation}
where \(k\) is the wave number. Substituting Equation
(\ref{psup7}) into Equation (\ref{claquan1}), we obtain:
\begin{align}
\widetilde{P}(k,t) =& \frac{e^{\frac{\hbar^2 \left(-(d m+\hbar k
t)^2-4 k^2 m^2 \sigma ^4\right)-16 m^3 \sigma ^4 v_0 \left(\hbar
k+m v_0\right)}{8 \hbar^2 m^2 \sigma
   ^2}}}{2 \sqrt{2 \pi } \left(1+e^{-\frac{d^2}{8 \sigma ^2}-\frac{2 m^2 \sigma ^2
   v_0^2}{\hbar^2}}\right)}\Big[ 1+ \exp \left(\frac{d \hbar k t}{2 m \sigma ^2}+\frac{4 k m \sigma ^2
   v_0}{\hbar}\right) \nonumber \\\label{claquan2}  &+ 2  \exp \left(\frac{d (d m+2 \hbar k t)}{8 m \sigma ^2}+\frac{2 m
   \sigma ^2 v_0 \left(\hbar k+m v_0\right)}{\hbar^2}\right) \, \cos \left(\frac{1}{2} k \left(d-2 t
   v_0\right)\right)\Big].
\end{align}

To recover the quantum probability density \(|\Psi(x,t)|^2\), the
inverse Fourier transform is applied: $|\Psi(x,t)|^2 =
\frac{1}{\sqrt{2 \pi}} \int_{-\infty}^{\infty} \widetilde{P}(k,t)
e^{i k x} \, dk$. Indeed, if we substitute Equation
(\ref{claquan2}) into this inverse transform formula and perform
the corresponding integral, we obtain the function defined in
Equation (\ref{psup7}).

To obtain the classical probability density, where no interference
pattern exists, we will employ a truncated version of the inverse
Fourier transform. Instead of performing the standard integral:
$\frac{1}{\sqrt{2 \pi}} \int_{-\infty}^{\infty} \widetilde{P}(k,t)
e^{i k x} \, dk$, we will restrict the integration to a finite
interval $[-k_0,k_0]$. Remarkably, as we will discover, the
integral $\frac{1}{\sqrt{2 \pi}} \int_{-k_0}^{k_0}
\widetilde{P}(k,t) e^{i k x} dk$ describes the classical
probability density for specific values of $k_0$.

Therefore, let us define the function $P(x,t,k_0)$ through the
integral:
\begin{equation}
\label{claquan3} P(x,t,k_0) =   \frac{1}{\sqrt{2 \pi}}
\int_{-k_0}^{k_0} \widetilde{P}(k,t) e^{i k x} \, dk.
\end{equation}
Substituting Equation (\ref{claquan2}) into Equation
(\ref{claquan3}), we obtain:
\begin{align}
P(x,t,k_0) = & \frac{e^{\frac{m \left(\hbar \left(-3 d^2 m \sigma
^2-2 i d \hbar t x-12 m \sigma ^2 x^2\right)+8 m \sigma ^2 v_0
\left(d \hbar
   t-2 i m \sigma ^2 x\right)-12 \hbar m \sigma ^2 t^2 v_0^2\right)}{2 \hbar^3 t^2+8 \hbar m^2 \sigma ^4}}}{4 \sqrt{\pi } \sqrt{2 \sigma ^2+\frac{\hbar^2
t^2}{2 m^2 \sigma ^2}} \left(1+e^{-\frac{d^2}{8 \sigma ^2}-\frac{2
m^2
   \sigma ^2 v_0^2}{\hbar^2}}\right)} \Big[ \nonumber \\ & \;\;\;\;
   \big(\text{erf}(\beta_2)-\text{erf}(\beta_1)\big) e^{\frac{m^2 \sigma ^2 \left(d^2-2 d t v_0+4 t^2 v_0^2+4
   x^2\right)}{\hbar^2 t^2+4 m^2 \sigma ^4}} \nonumber \\ & + \big(\text{erf}(\beta_4)-\text{erf}(\beta_3)\big) e^{\frac{m \left(\hbar m \sigma ^2 \left(d^2-2 d t v_0+4 t^2 v_0^2+4
   x^2\right)+2 i d \hbar^2 t x+16 i m^2 \sigma ^4 v_0 x\right)}{\hbar^3 t^2+4 \hbar m^2 \sigma
   ^4}} \nonumber \\ &+ \big(\text{erf}(\gamma_2)-\text{erf}(\gamma_1)\big) e^{\frac{m \left(\hbar m \sigma ^2 \left(d^2-2 t v_0 (d+2 x)+2 d x+4
   t^2 v_0^2+4 x^2\right)+i d \hbar^2 t x+8 i m^2 \sigma ^4 v_0 x\right)}{\hbar^3 t^2+4 \hbar m^2 \sigma
   ^4}} \nonumber \\ \label{claquan4} &+ \big(\text{erf}(\gamma_4)-\text{erf}(\gamma_3)\big) e^{\frac{m \left(\hbar m \sigma ^2 \left(d^2-2 d \left(t
   v_0+x\right)+4 \left(t^2 v_0^2+t v_0 x+x^2\right)\right)+i d \hbar^2 t x+8 i m^2 \sigma ^4 v_0 x\right)}{\hbar^3 t^2+4 \hbar m^2
   \sigma ^4}} \Big].
\end{align}
The error function is denoted by erf in this equation. To simplify
the notation, we have defined the following values for $\beta_i$
and $\gamma_i$:
\begin{align}
\beta_1 &= \frac{h^2 t \left(d m-\hbar k_0 t\right)+4 m^2 \sigma
^4 \left(2 m v_0-\hbar k_0\right)-4 i \hbar m^2 \sigma ^2 x}{2
\hbar m \sigma
   \sqrt{2 \hbar^2 t^2+8 m^2 \sigma ^4}}, \\
\beta_2 &= \frac{h^2 t \left(d m+\hbar k_0 t\right)+4 m^2 \sigma
^4 \left(\hbar k_0+2 m v_0\right)-4 i \hbar m^2 \sigma ^2 x}{2
\hbar m \sigma
   \sqrt{2 \hbar^2 t^2+8 m^2 \sigma ^4}}, \\
\beta_3 &= \frac{\hbar^2 t \left(d m-\hbar k_0 t\right)+4 m^2
\sigma ^4 \left(2 m v_0-\hbar k_0\right)+4 i \hbar m^2 \sigma ^2
x}{2 \hbar m \sigma
   \sqrt{2 \hbar^2 t^2+8 m^2 \sigma ^4}}, \\
\beta_4 &=  \frac{\hbar^2 t \left(d m+\hbar k_0 t\right)+4 m^2
\sigma ^4 \left(\hbar k_0+2 m v_0\right)+4 i \hbar m^2 \sigma ^2
x}{2 \hbar m \sigma
   \sqrt{2 \hbar^2 t^2+8 m^2 \sigma ^4}}, \\
   \gamma_1 &= \frac{k_0 \left(-\hbar^2 t^2-4 m^2 \sigma ^4\right)-2 i m^2 \sigma ^2 \left(d-2 t v_0-2 x\right)}{2 m \sigma  \sqrt{2 \hbar^2
   t^2+8 m^2 \sigma ^4}}, \\
\gamma_2 &=\frac{k_0 \left(\hbar^2 t^2+4 m^2 \sigma ^4\right)-2 i
m^2 \sigma ^2 \left(d-2 t v_0-2 x\right)}{2 m \sigma  \sqrt{2
\hbar^2
   t^2+8 m^2 \sigma ^4}}, \\
\gamma_3 &= \frac{k_0 \left(-\hbar^2 t^2-4 m^2 \sigma ^4\right)+2
i m^2 \sigma ^2 \left(d-2 t v_0+2 x\right)}{2 m \sigma  \sqrt{2
\hbar^2
   t^2+8 m^2 \sigma ^4}}, \\
\gamma_4 &=  \frac{k_0 \left(\hbar^2 t^2+4 m^2 \sigma ^4\right)+2
i m^2 \sigma ^2 \left(d-2 t v_0+2 x\right)}{2 m \sigma  \sqrt{2
\hbar^2
   t^2+8 m^2 \sigma ^4}}.
\end{align}

\begin{figure}[h] \centering
    \includegraphics[width=\linewidth,height=115mm]{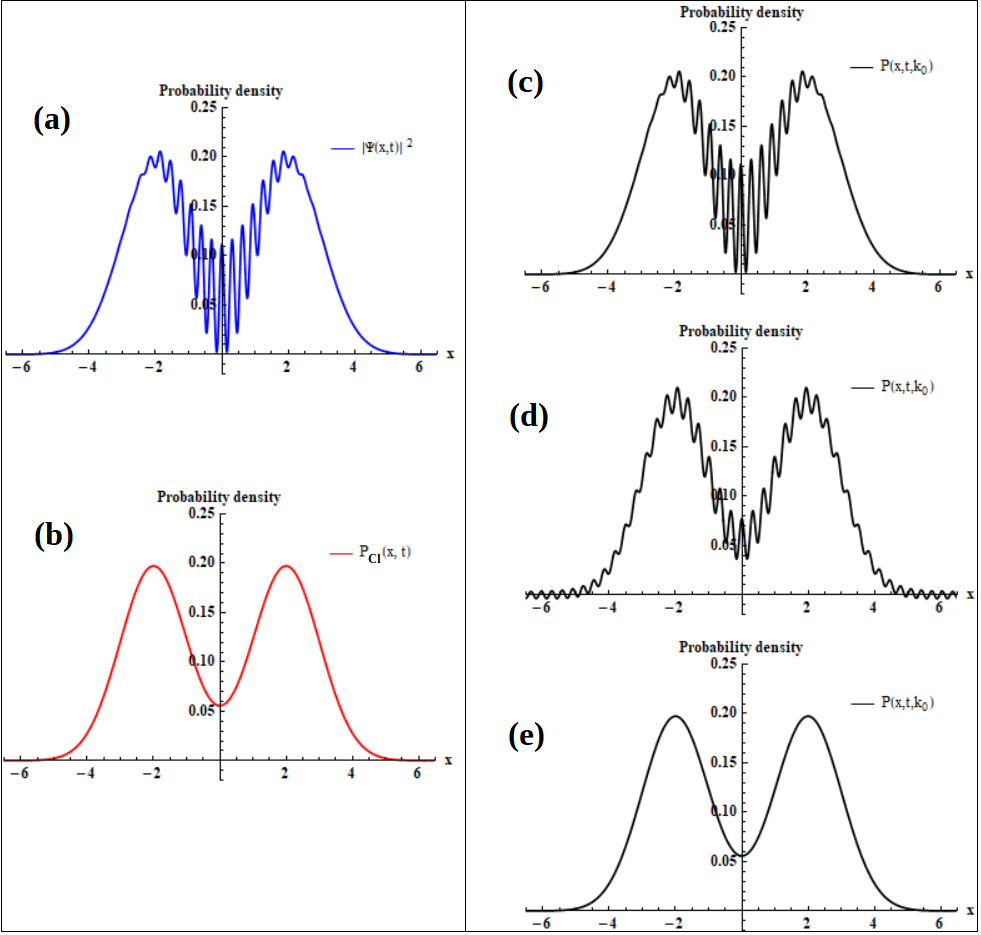}
    \caption{The following graphs depict curves defined by the functions $P_{\text{Cl}}(x, t)$ in red, $|\Psi(x,t)|^2$ in blue, and $P(x,t,k_0)$ in black.
    These curves were plotted using the parameter values $\hbar = \sigma = m = 1$, and $d = v_0 = 10$ at the time $t=0.3$. (a) shows
    the plot of the quantum probability density $|\Psi(x,t)|^2$. (b) shows the plot of the classical distribution $P_{\text{Cl}}(x, t)$. (c), (d),
    and (e) show a sequence of plots of the function $P(x,t,k_0)$ for the values of the parameter $k_0=30$, $k_0=20$, and $k_0=10$ respectively.}
    \label{fig3}
\end{figure}

From this point, we will refer to $P(x,t,k_0)$ as the truncated
quantum density. Since the expression for this function is
somewhat complicated, a general analytical analysis would be very
complex. Therefore, instead of such an analysis, we will present
the behavior of the truncated density for varying values of the
parameter $k_0$, using the parameter values $\hbar = \sigma = m =
1$, and $d = v_0 = 10$ at time $t=0.3$. For example, in Figure
\ref{fig3}, we have plotted the functions $P_{\text{Cl}}(x, t)$,
$|\Psi(x,t)|^2$, and $P(x,t,k_0)$.

In Figure \ref{fig3}, we observe an interesting behavior: by
varying the parameter \( k_0 \) in the truncated quantum density
\( P(x,t,k_0) \), the graph transitions from a quantum-like
distribution to a classical-like distribution. For values of \(
k_0 \) in the range \( k_0 \in (3,18) \), the graph of
$P(x,t,k_0)$ closely resembles the classical distribution \(
P_{\text{Cl}}(x, t) \), while for \( k_0 > 18 \), it converges
towards the quantum density \( |\Psi(x,t)|^2 \). The transition
between these regimes is relatively rapid, indicating a noticeable
change in behavior.

\begin{figure}[h]
    \centering
    \includegraphics[width=\linewidth,height=118mm]{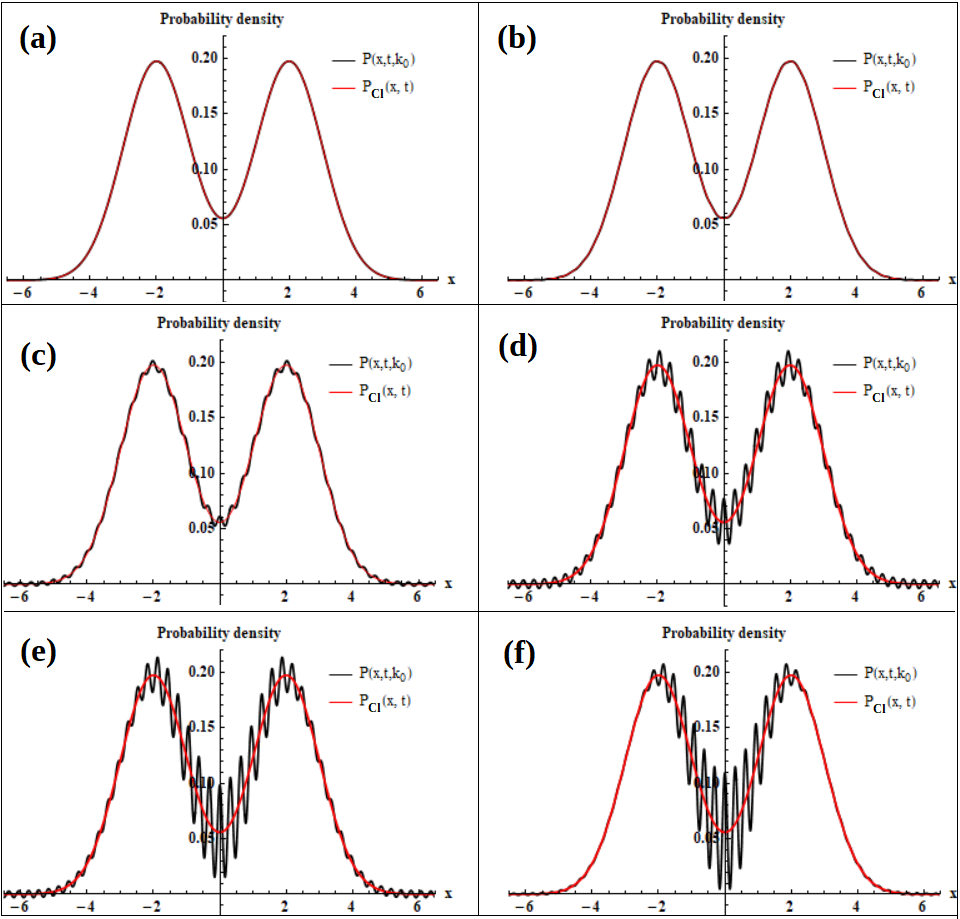}
    \caption{The following graphs depict curves defined by the functions \( P_{\text{Cl}}(x, t) \) in red, and \( P(x,t,k_0) \)
    in black. These curves were plotted using the parameter values \( \hbar = \sigma = m = 1 \), and \( d = v_0 = 10 \)
    at the time \( t = 0.3 \). The plots are presented in the order (a) \( k_0 = 17 \), (b) \( k_0 = 18 \), (c) \( k_0 = 19 \),
    (d) \( k_0 = 20 \), (e) \( k_0 = 21 \), and (f) \( k_0 = 22 \).}
    \label{fig4}
\end{figure}

To further illustrate this transition, we consider the same
parameter values \( \hbar = \sigma = m = 1 \), and \( d = v_0 = 10
\) at time \( t = 0.3 \). We study how the graph that represents
the function \( P(x,t,k_0) \) transitions to the graph of the
classical distribution \( P_{\text{Cl}}(x, t) \) for values close
to \( k_0 = 18 \). This result is presented in Figure \ref{fig4},
where it is observed that for values \( k_0 = 17 \) and \( k_0 =
18 \), the red curve representing the classical distribution given
by the function \( P_{\text{Cl}}(x, t) \) is practically
superimposed on the black curve representing the function \(
P(x,t,k_0) \). From the value \( k_0 = 18 \) onwards, we observe
that the curves begin to differentiate, with a noticeable
difference for values \( k_0 > 20 \).

To better visualize this transition at points close to $k_0=18$,
let us fix \( x = 0.5 \) and \( t = 0.3 \). Thus, we consider \(
|\Psi(0.5,0.3)|^2 \), \( P_{\text{Cl}}(0.5, 0.3) \), and \(
P(0.5,0.3,k_0) \) with the parameters having the values \( \hbar =
\sigma = m = 1 \), and \( d = v_0 = 10 \).

Here, \( |\Psi(0.5,0.3)|^2 \) and \( P_{\text{Cl}}(0.5, 0.3) \)
yield the fixed values $0.0378736$ and $0.0749317$ respectively,
while \( P(0,0.3,k_0) \) provides values that depend on \( k_0 \).

\begin{figure}[h]
    \centering
    \includegraphics[width=\linewidth]{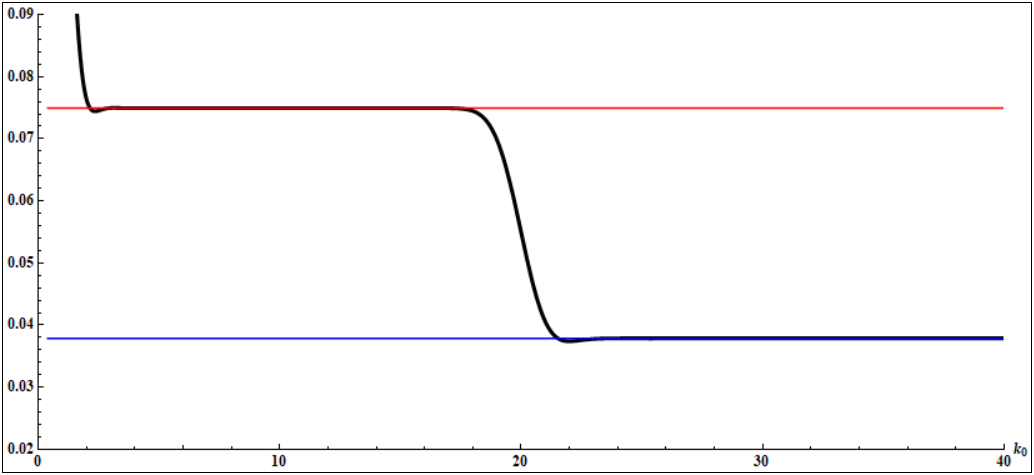}
    \caption{Transition from quantum to classical behavior as a function of \( k_0 \). The plot shows the
    value \( |\Psi(0.5,0.3)|^2 = 0.0378736 \) as the blue horizontal line, \( P_{\text{Cl}}(0.5, 0.3) = 0.0749317 \) as the red horizontal
    line, and \( P(0.5,0.3,k_0) \) represented as the black curve as \( k_0 \) varies.}
    \label{fig5}
\end{figure}

By plotting \( |\Psi(0.5,0.3)|^2 \), \( P_{\text{Cl}}(0.5, 0.3)
\), and \( P(0.5,0.3,k_0) \) on the vertical axis against \( k_0
\) on the horizontal axis, for \( k_0 \in (0, 40) \), we can
observe how \( P(0.5,0.3,k_0) \) transitions from the quantum
value \( |\Psi(0.5,0.3)|^2 = 0.0378736 \) to the classical value
\( P_{\text{Cl}}(0.5, 0.3) = 0.0749317 \) at points close to \(
k_0 = 18 \). This transition is depicted in Figure \ref{fig5}.

These results significantly enhance our understanding of the
emergence of classical behavior from quantum systems. By
meticulously analyzing the truncation of high wave numbers in the
Fourier transform of the quantum probability density, we gain
insight into how classical probabilities arise from quantum
mechanics.

In the subsequent section, we will delve into the same system
under investigation but from the perspective of decoherence
formalism. This serves to augment our current exploration,
leveraging the well-established principles of decoherence. It is
presented as a complementary avenue to demonstrate the
attainability of classical outcomes within our framework. This
forthcoming analysis will scrutinize in detail how the chosen
model aligns with decoherence theory.

\section{Superposition of Gaussian wavepackets and decoherence}
In this section, we revisit the previously studied quantum system,
which will be denoted by $\mathcal{S}$, interacting with its
environment, denoted by $\mathcal{E}$. We will explore how
decoherence arising from this interaction suppresses quantum
interference, leading to a classical probability distribution.

In the classical regime, the interpretation of the system
$\mathcal{S}$ is straightforward. Consider a particle of mass $m$
with its initial position drawn from the Gaussian distribution
discussed earlier. If the initial position is less than zero $(x <
0)$, the particle has a positive initial velocity $(v > 0)$.
Similarly, for an initial position greater than zero $(x > 0)$,
the particle has a negative initial velocity $(v < 0)$. The
distribution associated with these two possibilities is
represented by Gaussians centered at $x = -d/2$ and $x=d/2$. These
Gaussians then move towards each other; the Gaussian centered at
$x = -d/2$ moves to the right, and the Gaussian centered at
$x=d/2$ moves to the left. As time evolves, these two Gaussians
eventually overlap, and due to the classical nature of the system,
no interference pattern is observed.

In the quantum regime, we study the equivalent scenario using a
wave function, $\Psi(x, t)$, which represents the quantum state of
system $\mathcal{S}$. Here, to describe whether the particle
originates from the right side $(x > 0)$ or the left side $(x <
0)$, we can decompose the wave function $\Psi(x, t)$ into the sum
of two wave functions:
\begin{align}
\label{deco1} \Psi(x, t) &= \frac{1}{\sqrt{2 \pi \hbar}} \int_{-\infty}^{\infty}\widetilde{\psi}(p) e^{\frac{i p x}{\hbar}-
\frac{i p^2 t}{2 \hbar m}} dp=\Psi_L(x, t) + \Psi_R(x, t), \\
\label{deco2} \Psi_L(x, t) &= \frac{1}{\sqrt{2 \pi \hbar}} \int_{0}^{\infty}\widetilde{\psi}(p) e^{\frac{i p x}{\hbar}-\frac{i p^2 t}{2 \hbar m}} dp, \\
\label{deco3} \Psi_R(x, t) &= \frac{1}{\sqrt{2 \pi \hbar}}
\int_{-\infty}^{0}\widetilde{\psi}(p) e^{\frac{i p
x}{\hbar}-\frac{i p^2 t}{2 \hbar m}} dp,
\end{align}
where the function $\widetilde{\psi}(p)$ can be obtained by taking
the Fourier transform of the wave function $\Psi(x,0)$ given in
Equation (\ref{psup5}), so that
\begin{align}
\label{deco4} \widetilde{\psi}(p)=  \frac{1}{\sqrt[4]{2 \pi }}
\frac{\sqrt{\frac{\sigma }{\hbar}}}{\sqrt{1+e^{-\frac{d^2}{8
\sigma ^2}-\frac{2 m^2 \sigma ^2
   v_0^2}{\hbar^2}}}}\Big[ e^{-\frac{\sigma ^2 (p-m v_0)^2}{\hbar^2}+\frac{i d (p-m
   v_0)}{2 \hbar}}+e^{-\frac{\sigma ^2 (m
   v_0+p)^2}{\hbar^2}-\frac{i d (m v_0+p)}{2 \hbar}} \Big].
\end{align}

Since velocity is proportional to momentum $p$, the integral taken
over $p > 0$ corresponds to positive velocities. This signifies
that the particle (or wave in this quantum case) originates from
the left (L) side. Conversely, the integral over $p < 0$ signifies
negative velocities, indicating the particle (or wave) originates
from the right (R) side. Decomposing the wave function $\Psi(x,
t)$ into $\Psi_L(x, t)$ and $\Psi_R(x, t)$ allows us to analyze
its behavior based on the particle's direction of motion, similar
to the classical description with Gaussians moving in opposite
directions. This separation will also prove beneficial when
studying the system's interaction with the environment.

Substituting Equation (\ref{deco4}) into Equations (\ref{deco2})
and (\ref{deco3}), we explicitly obtain the left and right
components of the wave function:
\begin{align}
\label{deco5} \Psi_L(x, t) &= \frac{\sqrt{m \sigma }
e^{\frac{1}{16} d \left(\frac{d}{\sigma ^2}-\frac{8 i m
v_0}{\hbar}\right)} \Big(  e^{\alpha _1^2}
\left(1+\text{erf}\left(\alpha _1\right)\right)+e^{\alpha _2^2}
\left(1+\text{erf}\left(\alpha
   _2\right)\right) \Big)}{2 \sqrt[4]{2 \pi
   } \sqrt{\left(2 m \sigma ^2+i \hbar t\right) \left(e^{\frac{d^2}{8 \sigma ^2}+\frac{2 m^2 \sigma ^2
   v_0^2}{\hbar^2}}+1\right)}}, \\
\label{deco6} \Psi_R(x, t) &= \frac{\sqrt{m \sigma }
e^{\frac{1}{16} d \left(\frac{d}{\sigma ^2}-\frac{8 i m
v_0}{\hbar}\right)}\Big(  e^{\alpha _1^2}
\left(1-\text{erf}\left(\alpha _1\right)\right)+e^{\alpha _2^2}
\left(1-\text{erf}\left(\alpha
   _2\right)\right)\Big)}{2 \sqrt[4]{2 \pi }
   \sqrt{\left(2 m \sigma ^2+i \hbar t\right) \left(e^{\frac{d^2}{8 \sigma ^2}+\frac{2 m^2 \sigma ^2
   v_0^2}{\hbar^2}}+1\right)}},
\end{align}
where the quantities $\alpha_1$ and $\alpha_2$ are defined as
\begin{align}
\label{deco7} \alpha_1 = \frac{-4 m \sigma ^2 v_0+i \hbar (2
x-d)}{2 \hbar \sqrt{4 \sigma ^2+\frac{2 i \hbar t}{m}}}, \;\;\;\;
\alpha_2 = \frac{4 m \sigma ^2 v_0+i \hbar (d+2 x)}{2 \hbar
\sqrt{4 \sigma ^2+\frac{2 i \hbar t}{m}}}.
\end{align}

Initially, at \(t = 0\), the system interacts with the environment
\(\mathcal{E}\), and the total wave function of the system and
environment can be written as \(|\Psi_{\mathcal{SE}}\rangle =
|\Psi_0\rangle \otimes |\mathcal{E}_0\rangle\). Over time, the
interaction leads to the entanglement of the system states
\(|\Psi_L(t)\rangle\) and \(|\Psi_R(t)\rangle\) with different
states of the environment \(|\mathcal{E}_1(t)\rangle\) and
\(|\mathcal{E}_2(t)\rangle\), respectively. For \(t > 0\), the
total wave function is:
\begin{align}
\label{deco8} |\Psi_{\mathcal{SE}}(t)\rangle = |\Psi_L(t)\rangle
\otimes |\mathcal{E}_1(t)\rangle + |\Psi_R(t)\rangle \otimes
|\mathcal{E}_2(t)\rangle.
\end{align}
As we are going to see, it will be crucial that the environment
states \(|\mathcal{E}_1(t)\rangle\) and
\(|\mathcal{E}_2(t)\rangle\) become orthogonal for effective
decoherence.

Considering the density matrix of the total system
\(\rho_{\mathcal{SE}}(t) = \left| \Psi_{\mathcal{SE}}(t)
\right\rangle \left\langle \Psi_{\mathcal{SE}}(t) \right|\) and
taking the partial trace $\text{Tr}_\mathcal{E} \left( \left|
\Psi_{\mathcal{SE}}(t) \right\rangle \left\langle
\Psi_{\mathcal{SE}}(t) \right| \right)$ over the environment
\(\mathcal{E}\), we obtain:
\begin{align}
\rho_\mathcal{S}(t) &= \text{Tr}_\mathcal{E} \left( \left|
\Psi_{\mathcal{SE}}(t) \right\rangle \left\langle
\Psi_{\mathcal{SE}}(t) \right| \right) \nonumber \\ \label{deco9}
& = \left| \Psi_L(t) \right\rangle \left\langle \Psi_L(t) \right|
+ \left| \Psi_R(t) \right\rangle \left\langle \Psi_R(t) \right| +
\langle\mathcal{E}_1 | \mathcal{E}_2\rangle \left( \left|
\Psi_L(t) \right\rangle \left\langle \Psi_R(t) \right| + \left|
\Psi_R(t) \right\rangle \left\langle \Psi_L(t) \right| \right),
\end{align}
where \(\langle\mathcal{E}_1 | \mathcal{E}_2\rangle\) is the inner
product between the environment states
\(|\mathcal{E}_1(t)\rangle\) and \(|\mathcal{E}_2(t)\rangle\).

Since the term $\langle\mathcal{E}_1 | \mathcal{E}_2\rangle \left(
\left| \Psi_L(t) \right\rangle \left\langle \Psi_R(t) \right| +
\left| \Psi_R(t) \right\rangle \left\langle \Psi_L(t) \right|
\right)$ arises from the off-diagonal terms of the total density
matrix, cancelling these terms reduces quantum probability to
classical probability. This cancellation occurs when the
environment states are orthogonal, i.e., $\langle\mathcal{E}_1 |
\mathcal{E}_2\rangle=0$. However, note that the term $\left(
\left| \Psi_L(t) \right\rangle \left\langle \Psi_R(t) \right| +
\left| \Psi_R(t) \right\rangle \left\langle \Psi_L(t) \right|
\right)$ could also be zero, independent of $\langle\mathcal{E}_1
| \mathcal{E}_2\rangle$. Therefore, to justify the assumption that
$\langle\mathcal{E}_1 | \mathcal{E}_2\rangle= 0$, one would need
to demonstrate that $\left( \left| \Psi_L(t) \right\rangle
\left\langle \Psi_R(t) \right| + \left| \Psi_R(t) \right\rangle
\left\langle \Psi_L(t) \right| \right)$ does not vanish. This
demonstration can be explicitly shown since the functions
$\Psi_L(x,t)=\langle x | \Psi_L(t) \rangle$ and
$\Psi_R(x,t)=\langle x | \Psi_R(t) \rangle$ are known. The
quantity $\left( \left| \Psi_L(t) \right\rangle \left\langle
\Psi_R(t) \right| + \left| \Psi_R(t) \right\rangle \left\langle
\Psi_L(t) \right| \right)$, in the position basis states $\left| x
\right\rangle$, is equivalent to
$2$Re$(\Psi^*_L(x,t)\Psi_R(x,t))$. Thus, if we have that
$2$Re$(\Psi^*_L(x,t)\Psi_R(x,t)) \neq 0$ to reduce the quantum
probability to the classical probability, we must have
$\langle\mathcal{E}_1 | \mathcal{E}_2\rangle=0$. All these
important results can be explicitly tested in our studied system.
For this purpose, we need to calculate the density operator
(\ref{deco9}) in the position basis states $\left| x
\right\rangle$, namely:
\begin{align}
\label{deco10} \langle x | \rho_\mathcal{S}(t) | x \rangle =
|\Psi_L(x, t)|^2 + |\Psi_R(x, t)|^2 + 2 \langle\mathcal{E}_1 |
\mathcal{E}_2\rangle \text{Re}(\Psi^*_L(x,t)\Psi_R(x,t)).
\end{align}
The quantity on the right hand side of Equation (\ref{deco10}),
which is the sum of the squares of the left and right components
of the wave function given in Equations (\ref{deco5}) and
(\ref{deco6}), is expected to represent the classical probability
density given in Equation (\ref{psup4}).

To visualize the behavior of the classical and quantum probability
densities over time with the parameters \( \hbar = \sigma = m = 1
\) and \( d = v_0 = 10 \), we plotted the following quantities:
the classical probability density \( P_{\text{Cl}}(x, t) \), the
sum of the squared magnitudes \( |\Psi_L(x, t)|^2 + |\Psi_R(x,
t)|^2 \), and the interference term \( 2
\text{Re}(\Psi^*_L(x,t)\Psi_R(x,t)) \). We considered a sequence
of times \( t = 0.2, 0.3, \) and \( 0.4 \).
\begin{figure}[h]
    \centering
    \includegraphics[width=\linewidth]{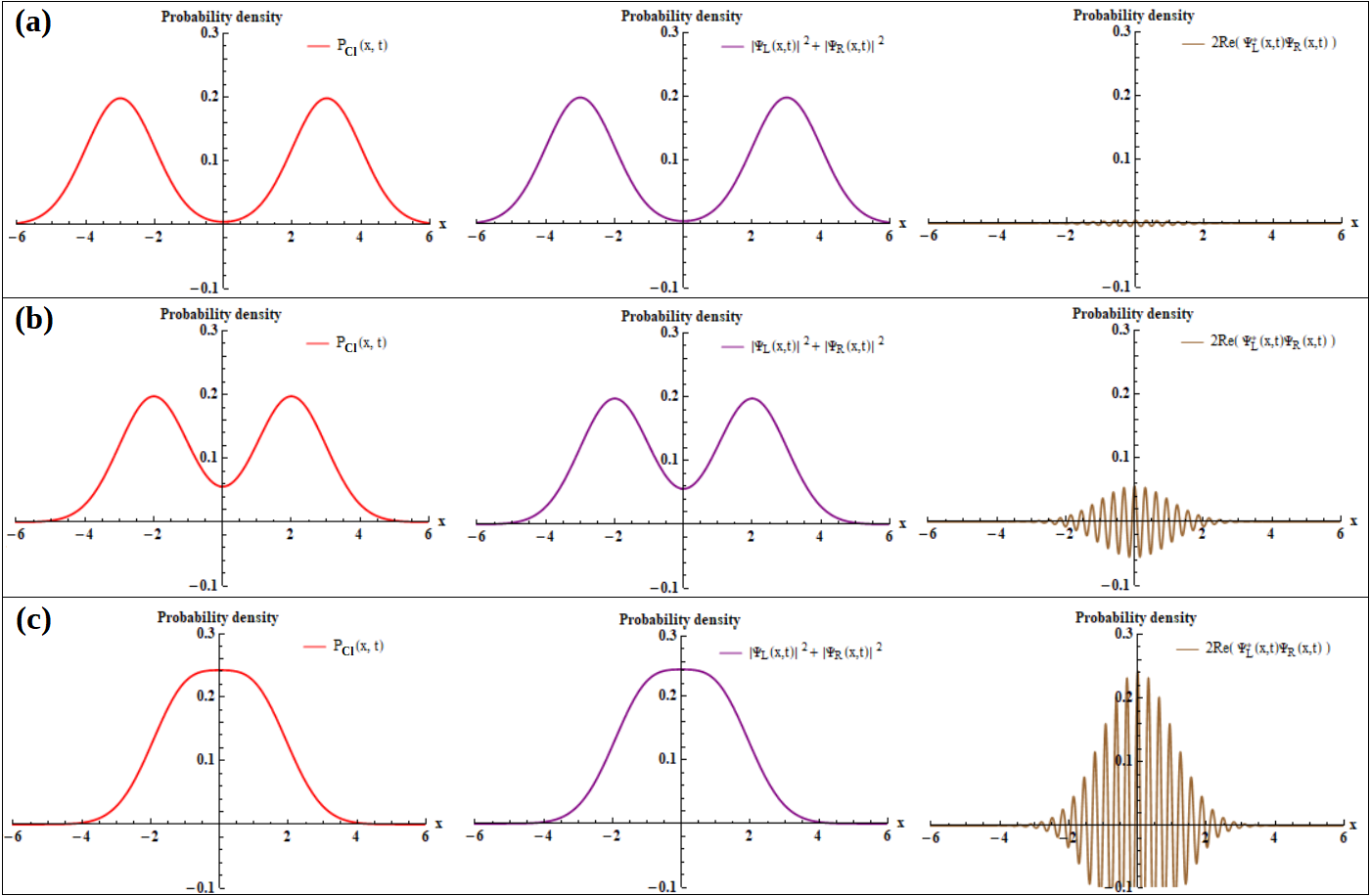}
    \caption{Plots of the classical probability density \( P_{\text{Cl}}(x, t) \) in red, the sum of the squared magnitudes
\( |\Psi_L(x, t)|^2 + |\Psi_R(x, t)|^2 \) in purple, and the
interference term \( 2 \text{Re}(\Psi^*_L(x,t)\Psi_R(x,t)) \) in
brown for different times: (a) \( t = 0.2 \), (b) \( t = 0.3 \),
and (c) \( t = 0.4 \). The classical probability density \(
P_{\text{Cl}}(x, t) \) closely matches \( |\Psi_L(x, t)|^2 +
|\Psi_R(x, t)|^2 \), indicating that \( 2
\text{Re}(\Psi^*_L(x,t)\Psi_R(x,t)) \) accounts for the
interference effects.}
    \label{fig6}
\end{figure}

As shown in Figure \ref{fig6}, the classical probability density
\( P_{\text{Cl}}(x, t) \) closely matches the quantity \(
|\Psi_L(x, t)|^2 + |\Psi_R(x, t)|^2 \), indicating that the term
\( 2 \text{Re}(\Psi^*_L(x,t)\Psi_R(x,t)) \) is responsible for the
interference effects. Recall that this interference term is
multiplied by the factor \( \langle \mathcal{E}_1 | \mathcal{E}_2
\rangle \), which implies that to eliminate the interference term,
\( \langle \mathcal{E}_1 | \mathcal{E}_2 \rangle \) needs to
vanish.

Another observation from the plots is that at times close to \( t
= 0 \), the interference term \( 2
\text{Re}(\Psi^*_L(x,t)\Psi_R(x,t)) \) is negligible, so initially
it is not necessary that \( \langle \mathcal{E}_1 | \mathcal{E}_2
\rangle = 0 \). In fact, since at \( t = 0 \), the system is not
yet entangled with the environment, the environmental states \(
\mathcal{E}_1 \) and \( \mathcal{E}_2 \) both approximate the
initial state \( \mathcal{E}_0 \). Thus, for \( t \to 0 \), we
have \( \mathcal{E}_1 \to \mathcal{E}_0 \) and \( \mathcal{E}_2
\to \mathcal{E}_0 \), leading to \( \langle \mathcal{E}_1 |
\mathcal{E}_2 \rangle \to \langle \mathcal{E}_0 | \mathcal{E}_0
\rangle = 1 \). However, as time progresses, the interference term
\( 2 \text{Re}(\Psi^*_L(x,t)\Psi_R(x,t)) \) starts to deviate from
zero. To eliminate this interference term for \( t > 0 \), \(
\langle \mathcal{E}_1 | \mathcal{E}_2 \rangle \) needs to vanish.

\section{Summary and discussion}
In conclusion, we have investigated the transition from quantum to
classical behavior using the one-dimensional free particle system.
Our findings show that under certain conditions, the quantum
probability density, derived from the Schr\"{o}dinger equation for
a Gaussian initial wave function, is equivalent to the classical
distribution of final positions obtained by convolving Gaussian
distributions of initial positions and velocities. We have also
demonstrated that when the initial state is a superposition of
Gaussian wave packets, quantum interference leads to a divergence
between classical and quantum descriptions. In such cases, we have
proposed a novel method to recover the classical distribution from
the quantum one using truncated Fourier transform analysis.

In our specific example, we observe that by applying Fourier
analysis to the quantum probability density and considering a
finite range of wavenumbers, we can recover the classical
probability distribution. The physical interpretation of this
method is that the high-frequency components of the quantum
probability density correspond to rapid oscillations associated
with quantum interference. By truncating the range of the
wavenumbers, we are essentially averaging out these oscillations,
leading to a smoother, classical-like distribution.

While this approach provides an alternative mathematical pathway
to recover classical behavior, it is conceptually related to
decoherence. Decoherence, a fundamental process in quantum
mechanics, explains the transition from the quantum to the
classical world by considering the interaction of a quantum system
with its environment. The suppression of high-frequency components
in our method can be seen as a manifestation of decoherence, where
the environment plays the role of measuring or observing the
system, leading to the emergence of classical behavior. Thus,
while our method is a mathematical procedure, it is grounded in
the physical reality that the classical limit emerges when quantum
coherence is reduced.

This method can be generalized beyond the specific example
presented here. The key idea is that the quantum-to-classical
transition involves filtering out components that contribute to
interference. Whether through environmental interactions
(decoherence) or mathematical truncation (Fourier analysis), the
result is a classical probability distribution. Our work
contributes to the understanding of the quantum-to-classical
transition and its physical foundations. These results have
important implications in various fields, from fundamental physics
to emerging quantum technologies. We hope that this study will
inspire future research in this fascinating area.




\bibliographystyle{ieeetr} 
\bibliography{bibcylib} 
\end{document}